\documentclass[lnbip]{svmultln}

\usepackage{makeidx}  
\usepackage{graphicx}
\graphicspath{{images/}}
\usepackage{todonotes}
\usepackage{hyperref} 

\usepackage{csquotes}

\begin{document}
\mainmatter              
\title{Towards an Architecture-centric Methodology for Migrating to Microservices}
\titlerunning{Microservices Migrations}  
\author{Jonas Fritzsch\inst{1} \and Justus Bogner\inst{1}
        \and Markus Haug\inst{1}
        \and Stefan Wagner\inst{1} \and Alfred Zimmermann\inst{2}}

\authorrunning{Jonas Fritzsch et al.}   
%
\tocauthor{Jonas Fritzsch, Justus Bogner, Stefan Wagner, Alfred Zimmermann}
\institute{
    University of Stuttgart, Stuttgart, Germany\\
    \email{{firstname.lastname}@iste.uni-stuttgart.de}\\
    \and
    University of Applied Sciences Reutlingen, Reutlingen, Germany\\
    \email{alfred.zimmermann@reutlingen-university.de}
}

\maketitle              

\begin{abstract}        
The euphoria around microservices has decreased over the years, but the trend of modernizing legacy systems to this novel architectural style is unbroken to date. A variety of approaches have been proposed in academia and industry, aiming to structure and automate the often long-lasting and cost-intensive migration journey. However, our research shows that there is still a need for more systematic guidance. While grey literature is dominant for knowledge exchange among practitioners, academia has contributed a significant body of knowledge as well, catching up on its initial neglect. A vast number of studies on the topic yielded novel techniques, often backed by industry evaluations. However, practitioners hardly leverage these resources. In this paper, we report on our efforts to design an architecture-centric methodology for migrating to microservices. As its main contribution, a framework provides guidance for architects during the three phases of a migration. We refer to methods, techniques, and approaches based on a variety of scientific studies that have not been made available in a similarly comprehensible manner before. Through an accompanying tool to be developed, architects will be in a position to systematically plan their migration, make better informed decisions, and use the most appropriate techniques and tools to transition their systems to microservices.

\end{abstract}
\section{The Challenge of Moving to Microservices}
In times of cloud-based software solutions, the microservices architectural style has become the de facto standard for large-scale and cloud-native commercial applications \cite{Vale2022}. Technological advancements like containerization and automation have paved the way for efficiently operating almost any number of independent functional units. However, existing legacy systems are often designed as monoliths and can therefore barely benefit from advantages such as improved scalability, maintainability, and agility through independent deployment units \cite{Jamshidi2018}. Hence, many companies try to migrate their systems towards microservices. While a rewrite of the entire application is expensive and often infeasible, architects are looking for less resource-intensive approaches to modernize a system. Architectural refactorings that are (partly) automated may reduce effort and risk of a migration tremendously. Unfortunately, there is no general approach that fits for arbitrary systems \cite{Fritzsch2019a}. A thorough analysis is required to choose the appropriate strategy and refactoring technique. 

Early adopters of microservices had to deal with manifold challenges, such as a lack of technical guidance and best practices, immature tooling, or organizational aspects. While pioneers like Amazon, Netflix, Spotify, and even the German retailer Otto published on their journeys of microservices adoption, such exemplary cases do not necessarily qualify as a blueprint. The average enterprise system has often more sophisticated tasks to fulfill than the well-studied retail domain, and moreover, IT budgets are often tight. Strict compliance and high-quality requirements do not leave much room for experimentation and failed investments. Hence, architects often struggle to find suitable guidance on planning and conducting such an architecture change systematically. 
In the following two subsections, we briefly summarize our view on the typical progress of a system migration and describe the gap between academia and industry.

\subsection{Three Phases of a Migration} \label{Three_Phases_of_a_Migration}

The initiation of a migration is commonly associated with the definition of strategic goals.
Subsequently, quality requirements are determined by all stakeholders of the systems. They serve as a measure for assessing the legacy system and potential alternatives. Hence, the outcome of this first phase should be a grounded decision for or against a modernization, based on specific quality attributes and metrics. While monolith and microservices are not the only possible architectural styles, we exclusively focus on the contraposition of these two patterns. 

Given that the first phase resulted in favor of a migration, phase two aims at defining an adequate migration strategy. 
One of the two major tasks in this phase is the definition of a development process based on different types of software modernization \cite{Bajaj2021}. 
Distinguishing greenfield and brownfield developments, we further split up the second in a re-build or re-factor development type. Again, our focus in this paper is on brownfield developments. 
Further distinctions can be made that affect the timeframe and consumption of resources in a migration process. While a \textit{big bang} migration aims to minimize the duration, a continuous evolution approach tries to minimize the needed resources.
The other major task in phase two is the selection of a service identification approach. 
Applied to the existing system, it will yield a suitable service cut and thereby determine the granularity of services. Our previous research has shown that deciding on the decomposition is often a manual task \cite{Fritzsch2019d} guided by methods like Domain-Driven Design (DDD).
However, a variety of existing artifacts from the legacy system can be beneficial for automating this task to some extent, e.g., code bases, databases, version control system data, logs and traces, and various other design documents or models. 
They provide valuable input for generating alternative decompositions in a more efficient way. 

After the migration has passed system comprehension and planning phases, the actual implementation phase starts. 
The elaborated strategy implies boundaries for duration, required resources, as well as needed organizational changes. The latter are in particular relevant as a consequence of altering a system's architecture \cite{Conway2018}.
Microservices candidates and target architecture are defined first, based on the approach and techniques chosen in phase two. The following implementation of services commonly iterates through several cycles. A cycle is characterized by the implementation of one or more microservices and a subsequent quality assessment of the emerging target system. The organizational changes, infrastructure build-up, and establishment of DevOps processes go hand in hand with the growing size of the system. 
That way, inadequacies of the architecture definition or even an unsuitable service identification approach can be corrected early with reasonable effort. 

\subsection{The Academia-industry Gap} \label{The_Academia-industry_Gap}

A rapidly growing number of scientific publications deal with the topic of microservices migrations, as the meta studies by Schroer et al. \cite{Schroer2020} and Ponce et al. \cite{Ponce2019} show. Existing research covers a variety of topics, starting from decision-making 
over process strategies \cite{Bajaj2021} to quality assurance \cite{Cojocaru2019a} and organizational aspects \cite{Fritzsch2019d}. 
The challenging question of service identification techniques in general \cite{Abdellatif2021a} and for microservices specifically is targeted by several dozen studies \cite{Fritzsch2019a,Ponce2019}. 
However, our empirical research has shown that this extensive body of scientific literature is mostly unknown to practitioners and therefore rarely leveraged \cite{Fritzsch2019d}. We found that even specialized consultancy companies do rarely consider such knowledge. There may be several aspects to this barrier, e.g., reservations regarding scientific databases, access limitations, or concerns regarding the practical applicability and relevance of scientific research. 
Hence, a key motivation of our work is in filtering, pre-processing, and presenting the relevant works to practitioners based on their specific systems and migration scenarios.

\section{Research Design}

\begin{figure}[ht]
    \centering
    \includegraphics[keepaspectratio=true, width=1.0\textwidth]{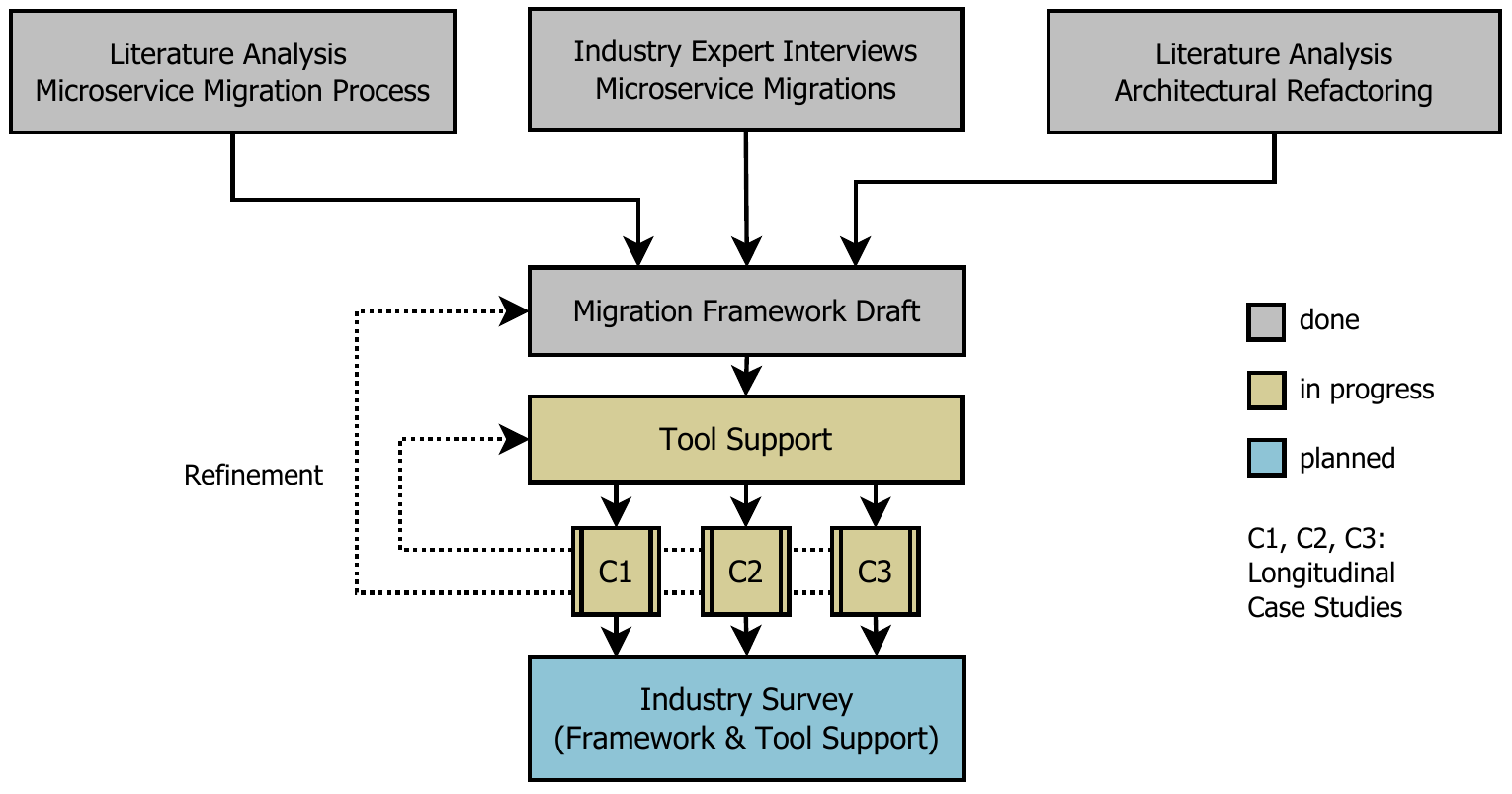}
    \caption{Research Method}
    \label{fig:MigrationFrameworkResearchMethod}
\end{figure}

Figure \ref{fig:MigrationFrameworkResearchMethod} illustrates our overall research method. The research objective is framed by the following questions:

\begin{enumerate}
    \itemsep 0em 
    \item How can a process framework represent a holistic view on microservice migration activities with a focus on architectural refactoring techniques?
    \item How can tool support based on such a framework provide guidance for architects in a specific migration scenario? 
\end{enumerate}

\noindent
As a foundation, we analyzed existing literature on the microservice migration process. In an interview study among 16 practitioners from 10 companies, we analyzed 14 systems from various domains regarding intentions, practices, and challenges \cite{Fritzsch2019d}. In addition, we contributed an early meta study classifying architectural refactoring approaches for migrations to microservices \cite{Fritzsch2019a}. 
 Our resulting methodology serves as a basis for three longitudinal case studies that are currently conducted in cooperation DATEV eG and Siemens AG. In an iterative process, the framework and accompanying tool support will be evaluated and refined. As a final step, we plan a large-scale survey among practitioners to assess the accompanying tools' applicability in certain contexts, its usefulness, and usability. 

\noindent
In the following section, we outline additional research foundations that we build upon and refer to related works. 

\section{Related Work}

According to the outlined research method, we split up the discussion of related studies into three clusters: 1) migration process, 2) architectural refactoring, and 3) associated aspects like quality assurance and re-organization.

\medskip
\noindent
\textbf{1) Migration Process} In their survey among 18 practitioners, Di Francesco et al. collected the various activities carried out in a migration to microservices \cite{DiFrancesco2018a}. 
The work provides an empirically collected, bottom-up classification of common activities. 

Taibi et al. followed a similar survey-based approach when querying 21 practitioners \cite{Taibi2017a}. They reconstructed a migration process framework that reflects the interviewees' procedures and best practices. In addition to Di Francesco et al., they also distinguish between re-development and continuous evolution, applying the popular Strangler pattern.

Wolfart et al. approach the migration topic more holistically in their work-in-progress paper \cite{Wolfart2021}. They analyzed six primary studies to come up with a unified process. 
In a related paper, Wolfart et al. propose a migration roadmap that builds upon this groundwork \cite{Wolfart2021a}. To this end, they conduct a more comprehensive systematic mapping of 62 primary studies dealing with the modernization of legacy systems to microservices. Their resulting framework depicts eight activities grouped into four phases, namely Initiation, Planning, Execution, and Monitoring. 

In the same way, we can regard the roadmap by Bozan et al. \cite{Bozan2021} on incrementally transitioning to a microservices architectures, which was distilled from interviews with 31 software experts. In contrast to the above-mentioned studies, the authors also reflect on the organizational and business-related impacts. 

\medskip
\noindent
\textbf{2) Architectural Refactoring} 
The secondary studies by Abdellatif et al.\cite{Abdellatif2021a}, Bajaj et al. \cite{Bajaj2021}, Schroer et al. \cite{Schroer2020}, Ponce et al. \cite{Ponce2019}, and Fritzsch et al. \cite{Fritzsch2019a} provide a holistic overview of this major technical challenge in a migration. Our earlier study \cite{Fritzsch2019a} attempted a classification of approaches based on their underlying techniques. Abdellatif et al. \cite{Abdellatif2021a} developed a more elaborate taxonomy that provides a solid foundation for use in our framework.
The majority of studies can be ascribed to at least one of the three basic categories of techniques: model-driven, static analysis, and dynamic analysis \cite{Ponce2019}. 
In addition, organizational structures or metadata like version control history \cite{Fritzsch2019a} can also provide valuable input.  

\medskip
\noindent
\textbf{3) Associated Aspects} 
As initially set strategic goals and subsequently identified quality attributes largely steer the architecture transformation, quality assurance needs a strong focus, as outlined by Shahin et al. \cite{Schroer2020}.
This aspect is reflected in some of the above suggested frameworks during the initial phase (requirements and strategic goals) or in the form of verification and validation activities. 
As a basis for assessing the relevance of different quality attributes for microservices in general \cite{Bogner2019} and in the context of a migration \cite{Fritzsch2019d}, we build upon our earlier empirical research. It also revealed that organizational changes and social aspects such as a mindset change can have considerable impact on the migration process. 

\medskip
\noindent
Significant advances have been made in detailing the process of a migration, as well as in elaborating techniques for decomposing monolithic systems. However, there is no holistic methodology available that combines both aspects. In addition, our  research revealed the lack of a vehicle for knowledge transfer into practice. 
We aim to address this issue by suggesting a methodology that presents a holistic view on microservice migration activities, with a focus on architectural refactoring. It is enriched with a systematic quality assessment and a high degree of process automation.
Furthermore, we seek to develop tool support that guides architects in a migration scenario, thereby allowing them to leverage the comprehensive body of scientific knowledge. 

\section{Proposed Migration Framework}

\begin{figure}[ht]
    \centering
    \includegraphics[keepaspectratio=true, width=1.0\textwidth]{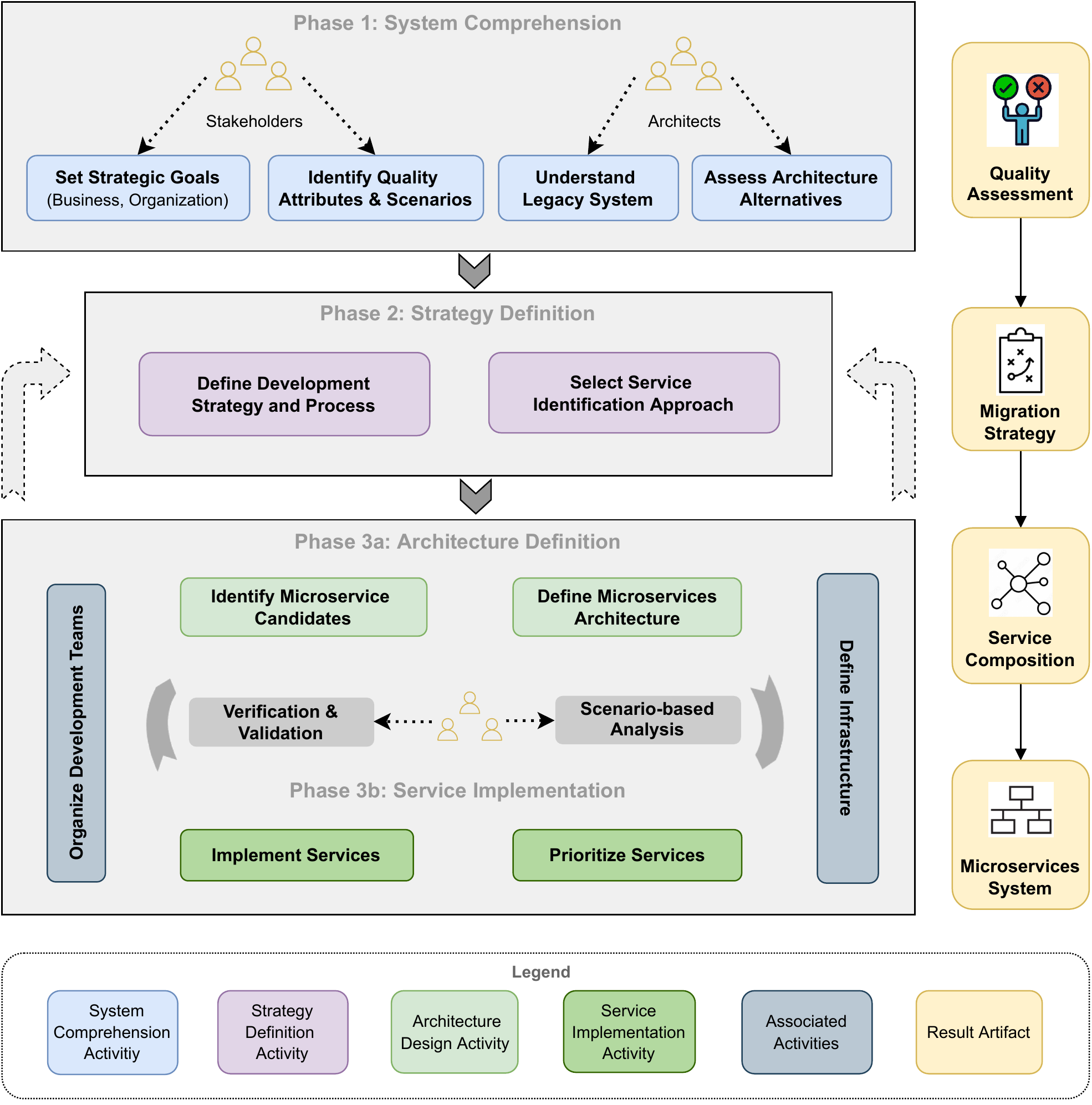}
    \caption{Proposed Framework for Microservices Migrations}
    \label{fig:MigrationFramework}
\end{figure}

Figure \ref{fig:MigrationFramework} shows our proposed architecture-centric framework for migrating to microservices. It incorporates ideas and groundwork of existing research, especially the works by Wolfart et al. \cite{Wolfart2021,Wolfart2021a}. Aspects of the works by Taibi et al. \cite{Taibi2017a} and Bozan et al. \cite{Bozan2021} have influenced the design as well. 
According to the discussion in \ref{Three_Phases_of_a_Migration}, we split a migration into three phases.
The reflected process may be applied separately for single subsystems as required.

\medskip
\noindent
\textbf{Phase 1} starts with a set of activities aiming to comprehend the existing system and assess alternatives as described by Wolfart et al. \cite{Wolfart2021,Wolfart2021a}. We depicted the common activities and involved personas. 
The activities in this phase are commonly performed as part of an architecture review using methods like ATAM, SAAM, or a more lightweight method, e.g., the one suggested by Auer et al. \cite{Auer2021}. 
The resulting quality assessment links the decision for or against a migration to microservices to distinct scenarios and associated quality requirements which the architectural styles in question are favorable for. 

\noindent
\textbf{Phase 2} entails the two activities described in Section~\ref{Three_Phases_of_a_Migration} to define the migration strategy. 
Depending on the system's technological state, organizational aspects, or other boundary conditions, different strategies may be chosen. 
The selection of a suitable approach and technique for service identification depends on several factors like targeted quality attributes, the available input artifacts, automation potential and maturity of available tool support.

\noindent
\textbf{Phase 3} starts with the identification of services and a preliminary definition of the target architecture. 
The incremental implementation of the identified services is preceded by a prioritization step. 
The framework puts a major focus on quality assurance aspects to ensure that initially defined measures are applied and satisfied. 
Hence, the implementation activities are accompanied by a scenario-based analysis and followed by a verification \& validation step. 
Deviations from the targets will consequently lead to altering the defined target architecture or even considering an alternative service identification approach by stepping back into phase 2.

\medskip
\noindent
As an overview of the migration progress, the activities' result artifacts are displayed in a flowchart on the right side next to each phase or group of activities.

\section{Current Status of Tool Support}

We recently conducted an interview study among software professionals on the framework's structure and automation capabilities.
The methodology is currently being refined and evaluated within longitudinal industry case studies. 
In parallel, we develop tool support in the form of a web-based application to provide guidance for architects. 
In phase 1, the tool collects system specifications and guides through the architecture assessment. 
This information serves as a documentation of the decision-making process and provides input for phase 2, which represents the tools' core functionality. 
Based on an extensible repository of existing approaches for service identification and architectural refactoring, the tool will assist in finding the appropriate technique for a specific system. 
The maintainable repository serves as a container for a growing list of approaches that have been published as peer-reviewed research papers. For their classification, we build upon the taxonomy proposed by Abdellatif et al. \cite{Abdellatif2021a} and our own preparatory work \cite{Fritzsch2019a}. 
The guidance provided by the tool in phase 3 will be realized as a selection of studies supporting the specific activities. 

We see potential for hosting the developed tool publicly and expanding it by functionality to incorporate user feedback or a rating system. 
In that regard, the framework and tool support could facilitate knowledge transfer not just from academia to industry but also vice versa, thereby contributing to close the gap highlighted in \ref{The_Academia-industry_Gap}.

%
%
\bibliographystyle{splncs03}
\bibliography{references}

\end{document}